\documentclass[prx,superscriptaddress,twocolumn,footinbib]{revtex4-1}
\pdfoutput=1
\usepackage[colorlinks=true,urlcolor=blue]{hyperref}
\usepackage{amsfonts}
\usepackage{graphicx}
\usepackage{placeins}
\usepackage{amsmath}
\usepackage{xcolor}
\usepackage{color}
\usepackage{bm}
\usepackage{ulem}

\definecolor{red}{rgb}{0.75,0,0}
\definecolor{blue}{rgb}{0,0,0.75}
\definecolor{green}{rgb}{0,0.5,0}

\begin{document}

\title{Defect-driven shape transitions in elastic active nematic shells}

\author{D. J. G. Pearce}
\affiliation{Dept.~of Theoretical Physics, University of Geneva, 1205 Geneva, Switzerland}
\affiliation{Dept.~of Biochemistry, University of Geneva, 1205 Geneva, Switzerland}
\affiliation{NCCR for Chemical Biology, University of Geneva, 1205 Geneva, Switzerland}
\affiliation{Dept.~of Mathematics, Massachusetts Institute of Technology, Cambridge, MA 02139, United States of America}
\author{S. Gat}
\affiliation{Dept.~of Chemical Engineering, Ben-Gurion University of the Negev, Beer Sheva 84105, Israel}
\author{G. Livne}
\affiliation{Dept.~of Chemical Engineering, Ben-Gurion University of the Negev, Beer Sheva 84105, Israel}
\author{A. Bernheim-Groswasser}
\email{bernheim@bgu.ac.il}
\affiliation{Dept.~of Chemical Engineering, Ben-Gurion University of the Negev, Beer Sheva 84105, Israel}
\affiliation{The Ilse Katz Institute for Meso and Nanoscale Science and Technology, Ben-Gurion University of the Negev, Beer Sheva 84105, Israel}
\author{K. Kruse}
\email{karsten.kruse@unige.ch}
\affiliation{Dept.~of Theoretical Physics, University of Geneva, 1205 Geneva, Switzerland}
\affiliation{Dept.~of Biochemistry, University of Geneva, 1205 Geneva, Switzerland}
\affiliation{NCCR for Chemical Biology, University of Geneva, 1205 Geneva, Switzerland}

\begin{abstract}

Active matter is characterized by its ability to induce motion by self-generated stress. In the case of a solid, such motion can lead to shape transformations. The stress-generating components can be anisotropic endowing the material with mesoscopic orientational order. It is currently unknown how the specific postions and orientations of these active constituents influence morphological changes. We study theoretically the effects of imposing topological point defects in the arrangements of the stress-generating components on the morphology of elastic active nematic shells. We show that topological defects of charge +1 are uniquely capable of increasing, reducing or maintaining the intrinsic curvature of the shell. These changes depend on the nature of the active stress and the phase angle of the defect. We apply our theory to experiments conducted on contracting actomyosin sheets. By combining defects of different charges, we can generate shells with arbitrary complexity. We confirm this flexibility by reproducing the shape of the freshwater polyp \textit{Hydra}, in which topological defects have been associated with morphological features of the animal. In addition to understanding morphogenetic processes, these principles can be applied to the design of programmable active mechanical metamaterials that form the basis of autonomous soft robots. 

\end{abstract}

\maketitle

\section{Introduction}
Mechanical stress plays a crucial role in many morphogenetic processes during animal development. Prominent examples are gastrulation or neural tube formation. This stress can be generated by the actin cytoskeleton, a polymer network consisting of actin filaments, myosin molecular motors, and other proteins~\cite{Alberts:2008mo}. How self-generated `active' stress is organized to lead to shape changes remains largely unknown. 

Due to the filamentous nature of the actin assemblies, actomyosin gels can exhibit macroscopic orientational order, which can extend over tissue length-scales~\cite{Gruler:1999bt,Duclos:2014bs,MaroudasSacks:2020eh}. Orientational order fields often exhibit topological defects~\cite{deGennes:2002vq}, where the orientation cannot be well defined, for example at the center of a vortex. In active materials, topological defects appear spontaneously~\cite{Sanchez:2012sp} and can focus stress~\cite{BlanchMercader:2020wc,BlanchMercader:2020tr,Hoffmann:2021}. 

Topological defects have been proposed to play important roles in organizing stress during organismal development~\cite{Saw:2017gn,Kawaguchi:2017em,MaroudasSacks:2020eh,Guillamat:2022}. A particularly striking recent example is given by the freshwater polyp \textit{Hydra}, in which the locations of the mouth, foot, and tentacles, were shown to correlate with topological defects in the actin network present in the early stages of morphogenesis~\cite{MaroudasSacks:2020eh}. How shape changes depend on the various features of topological defects remains an unsolved problem.

Reconstituted systems provide a powerful means to understand how the actin cytoskeleton induces cellular processes~\cite{Liu:2009da}. In this approach, components of the cytoskeleton are extracted from cells, purified, and studied in well-controlled environments. Such systems have played an important role in understanding aspects of cell motility~\cite{Cameron:1999,Bernheim:2002,Dayel:2009,Siton:2011}, the propensity of actin networks to self-organize~\cite{Backouche:2006,Reymann:2012}, to contract~\cite{Kohler:2012,Ideses:2013fr,Alvarado:2013,Linsmeier:2016,Ennomani:2016,Schuppler:2016}, and to generate shape changes~\cite{Boukellal:2004,Ideses:2018dna}.

In this work, we explore the morphology of thin elastic active nematic shells featuring topological defects. We first study thin disks with a single topological defect and compare our results to reconstituted actomyosin gels. We then extend our analysis to topological defects on thin spherical shells. Finally, we show how this approach can be used to recover the complex morphology of \textit{Hydra} from the positions and charges of topological defects. 

\section{Results}
\subsection{An agent-based model for an anisotropic elastic active nematic material} 
Consider an isotropic elastic material with embedded active components capable of locally inducing uniaxial expansion or contraction. This is in contrast to previous works that have examined isotropic active stresses~\cite{Zakharov:2021a,Zakharov:2021b,Ideses:2013fr}. Such active components could be pairs of actin filaments linked by myosin motors, actomyosin bundles, or elongated cells. 

We capture the effects of the material's activity by a change of its reference state in which the elastic energy is minimal~\cite{MatozFernandez:2020cl}. This is evidently appropriate if the material is a network of bundles linked at crossing sites and where the bundle lengths change or the crossing sites move due to active processes. However, this method applies to a broader class of active materials and the changing reference state is complementary to introducing an active stress~\cite{Berthoumieux:2014eo,Salbreux:2017ci,Morris:2019bx}.

We describe the material as a continuum and the (coarse-grained) anisotropy of the local expansions or contractions is captured by an orientation field $\underline{\hat{p}}$ with $\underline{\hat{p}}^2=1$. It is coupled to an order parameter, $S\in[0,1]$, accounting for the local degree of anisotropy. Let us point out that since the orientational order is nematic in character, the sign of $\underline{\hat{p}}$ is inconsequential to the mechanics of the system.

To analyze deformations due to changes in the reference state, we introduce an agent-based model. In this model, we partition the volume of the material into polyhedra by means of a Voronoi tessellation, see SI for details. The tessellation is such that there are multiple Voronoi cells across the shortest dimensions of the material, which ensures resistance to bending. The corresponding Delaunay triangulation describes a network connecting the centers of each Voronoi cell. We account for the average stress through the faces of the Voronoi cells by springs along the edges of the Delaunay network. It should be noted that the Dalaunay network is not a representation of the material's molecular structure.  

We describe changes of the reference state due to activity through time-dependent modifications of the springs' rest lengths. Explicitly, the rest length $l_i$ of spring $i$ at time $t$ with initial orientation $\underline{\hat{l}}_i$ and midpoint position $\underline{r}_i$ is given by
\begin{multline}
\label{eq:l}
l_i = \tilde{l}_i(1+\zeta(t))\\\times\sqrt{1 +\xi(t) S(\underline{r}_i) +\lambda(t) S(\underline{r}_i)[(\underline{\hat{p}}\cdot \underline{\hat{l}}_i)^2-0.5]},
\end{multline}
where $\tilde{l}_i$ is the initial rest length. The phenomenological parameters $\zeta$, $\xi$ and $\lambda$ represent active strain coefficients. Depending on their sign, they describe activity-induced expansion or contraction. The time-dependence of the strain coefficients reflects the evolution of the reference state and is, in principle, determined by dynamic equations capturing the effects of activity. Since we focus on how stress patterns induced by topological defects affect deformations of elastic active nematic materials, we refrain from giving these equations and instead prescribe in the following the coefficients' evolution. 

The differential anisotropic strain coefficient $\lambda$ controls changes of the rest length depending on their alignment with the orientation field, hence induces anisotropic contraction or expansion. We consider linear growth up to a predetermined stall value $\lambda^s$ at time $t^s$:
\begin{equation}
\label{eq:growth}
\lambda(t) = \begin{cases}
\frac{\lambda^st}{t^s}, &\text{if $t<t^s$}.\\
\lambda^s, &\text{otherwise}.
\end{cases}
\end{equation}
The differential isotropic strain coefficient $\xi$ controls changes of the rest length depending on the local order parameter, $S$. This then describes isotropic expansion or contraction regardless of the orientation of the spring. We set $\xi=0$; an in depth discussion of the effect of $\xi$ is presented in the SI. Finally, the global strain coefficient $\zeta$ represents homogenous, isotropic expansion or contraction, thus only contributes to a global re-scaling of the system and we set $\zeta=0$ for the remainder of this work. 

We consider the case where the intrinsic dynamics of the orientation field are negligible. Consequently, the values $\underline{\hat{p}}\cdot \underline{\hat{l}}_i$ and $S(\underline{r}_i)$ are defined on the un-deformed material and remain constant in time. This is consistent with an active elastic, in which these fields deform with the material. We choose all springs to have the same spring constant, $k=1$, and evolve the system according to an over-damped Langevin Equation. All simulations run for at least $2t^s$ and the final configurations of the gels are given as a function of the coefficients' stall values. 

\subsection{Defects in elastic active nematic disks} 
We first consider originally flat, thin circular disks with thickness $h$. Neglecting variations across the thickness of the sheet, we describe the planar nematic texture as
\begin{equation}
\label{eq:p}
\underline{\hat{p}} = {\cos(\theta)\hat{\underline{e}}_r} + {\sin(\theta)\hat{\underline{e}}_\Phi},
\end{equation}
where we have employed the orthonormal cylindrical basis $(\hat{\underline{e}}_r,\hat{\underline{e}}_\Phi,\hat{\underline{e}}_z )$. The angle $\theta$ describes the orientation of the nematic director relative to the radial direction $\hat{\underline{e}}_r$.

\begin{figure}
	\centering
	\includegraphics[width=\columnwidth]{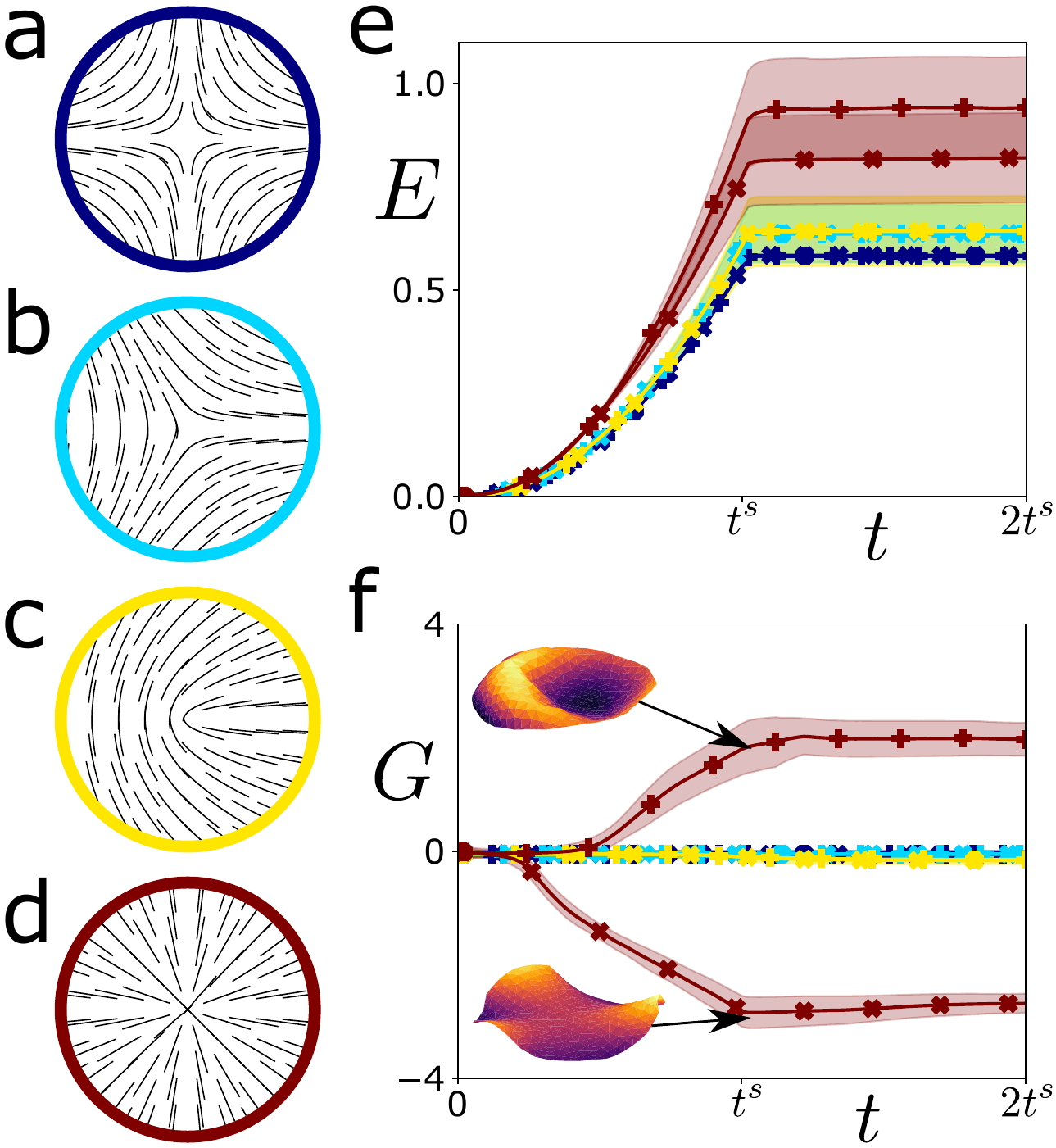}
	\caption{\label{fig:dyn} Dynamics of elastic active nematic disks. a-d) Nematic textures minimizing the Frank energy around topological defects with phase $\psi =0$ and charge $q=-1$ (a), $-1/2$ (b), $1/2$ (c), and $1$ (d). e) Elastic energy $E$ stored in and f) integrated Gaussian curvature $G$ of elastic active nematic disks as a function of time $t$. Colors in (e,f) correspond to those in (a-d). We used $\xi=0$, $\lambda^s=0.5$ ($\mathbf{+}$) and $-0.5$ ($\mathbf{\times}$).}
\end{figure}

We first consider a single topological defect with charge $q$ situated at the center of the material. The nematic texture around this defect is calculated by minimizing the Frank free energy to give
\begin{equation}
\label{eq:the}
\theta = (q-1)\Phi + \psi,
\end{equation}
where we have introduced the constant $\psi$ which is referred to as the phase of the defect. For a nematic, $q\in\frac{1}{2}\mathbb{Z}$, and the textures around the four lowest charge defects are shown in Fig.~\ref{fig:dyn}a-d. For all defects with $q\neq1$, changes of the phase are associated with a global rotation of the nematic texture around the defect. However, when $q=1$ the defect is an aster for $\psi=0$, a spiral for $0<|\psi|<\pi/2$, and a vortex for $\psi=\pi/2$, see SI. At the center of a topological defect, the nematic director is not well defined and the order parameter $S$ vanishes. Close to the center of the defect, it depends monotonically on the radial distance $r$. For simplicity we consider the case where $S(r) = r^2/R^2$ with $R$ being the initial radius of the disk. 

Distortions of an orientation order field typically come with an energy cost. In nematic solids, where orientation is tightly coupled to the solid's configuration, this can lead to interesting forms~\cite{Frank:2008,Modes:2011}. However, we consider the regime where the corresponding elastic constants associated with deformation of the nematic field are small compared to those associated with activity or deformation of the bulk material. This is consistent with the behavior of molecular motors within the cytoskeleton~\cite{Zhang:2018}.

The stored elastic energy in the material increases monotonically as the activity induces a change in the reference lengths of the springs, Fig.~\ref{fig:dyn}e. This energy cannot be dissipated since the strain pattern around the defect is incompatible with the disk shaped geometry of the material.  

In order to quantify the shapes of the disks, we introduce the integrated Gaussian curvature, $G = \int_A K $, which is dimensionless. We refer to the shapes with $G<0$ as `saddles' and with $G>0$ as `domes'. We see that only materials featuring a $q=1$ topological defect undergo a significant change in geometry with all other disks remaining essentially flat, Fig.~\ref{fig:dyn}f. This is particularly interesting, as previous studies have coupled Gaussian curvature to topological charge in active nematic systems~\cite{pearce2019geometrical,ellis2018curvature}. Furthermore, we see a delay in the adoption of positive Gaussian curvature relative to that of negative Gaussian curvature. This is due to a snap-through like instability in which sheets with $G>0$ globally break symmetry. This can lead to the sheets being trapped in a state with sub-optimal Gaussian curvature, Fig.~\ref{fig:dyn}f, inset. 

\begin{figure}
	\includegraphics[width=\columnwidth]{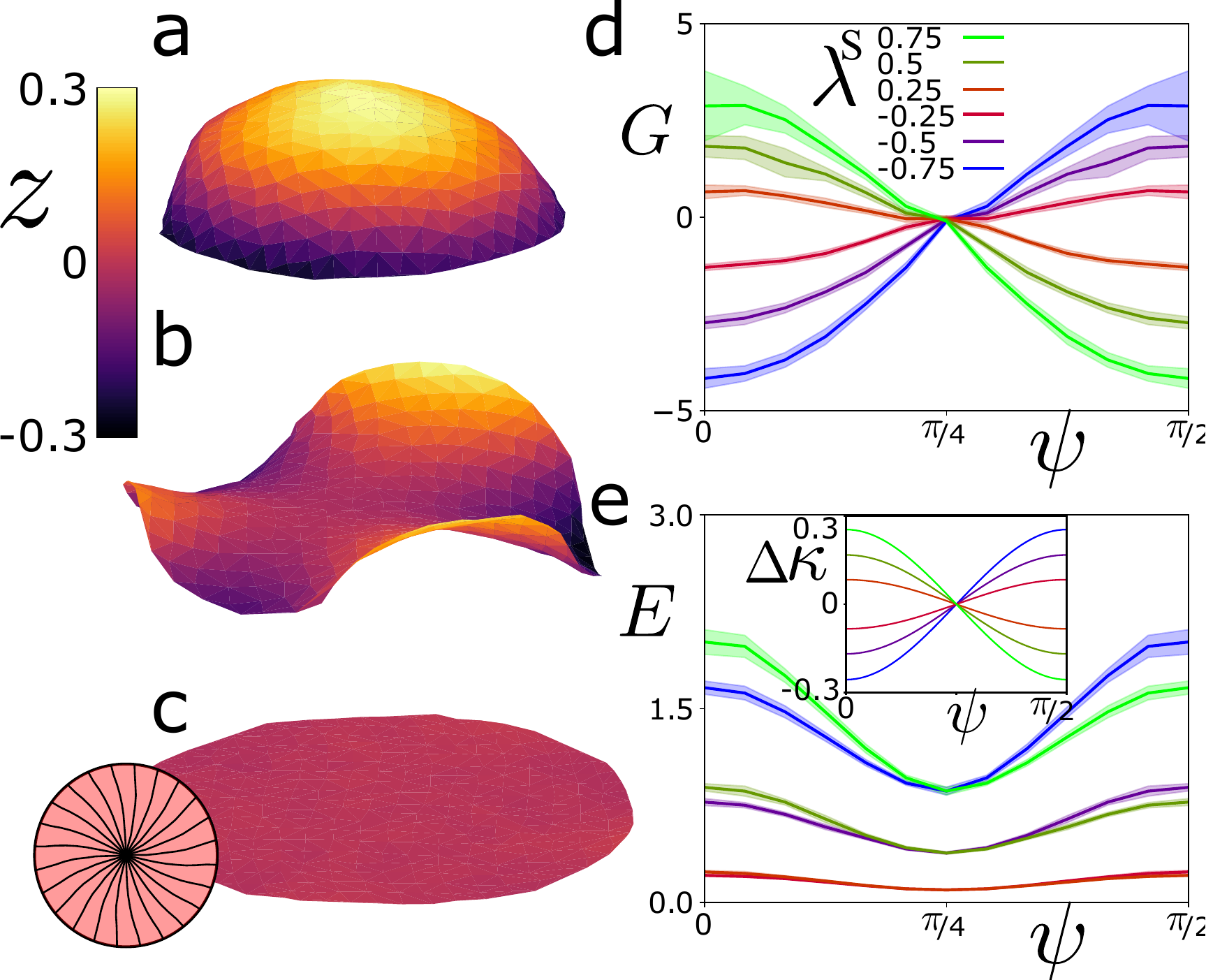}
	\caption{\label{fig:GA} Deformations around a topological defect featuring purely anisotropic active stress.  a-c) Final surfaces for $\lambda^s=0.5$ and $\psi = 0$  (a), $\psi = \pi/2$  (b), and $\psi = \pi/4$ (c). Inset: final position of initially straight lines. Color indicates height $z$. d) Integrated Gaussian curvature $G$, e) stored elastic energy as a function of the phase $\psi$ for various values of $\lambda^s$. Inset: $\Delta\kappa$ for the same range of $\psi$ and $\lambda^s$. }
\end{figure}

The sign of the Gaussian curvature induced around a topological defect with charge $q=1$ is the same as the sign of $\lambda^s$, Fig.~\ref{fig:dyn}f. This implies that such a  defect is capable of generating both, dome and saddle shapes, Fig.~\ref{fig:GA}a,b. By varying the phase $\psi$ and the sign of $\lambda^s$of the $q=1$ topological defect and the sign of the active stress, it is possible to generate a continuum of Gaussian curvatures, Fig.~\ref{fig:GA}d. The Gaussian curvature passes through zero at $\psi=\pi/4$ for all values of $\lambda^s$ at which the sheet remains approximately flat, Fig.~\ref{fig:GA}c,d. 

We find $G(\psi, \lambda)=G(\pi/2-\psi,-\lambda)$. This property is due to the fact that a phase change of $\Delta\psi=\pi/2$ is equivalent to changing the sign of $\lambda$, Eq.~\eqref{eq:l}. Once again we see that the magnitude of positive Gaussian curvature is lower than the equivalent negative Gaussian curvature when $\lambda\rightarrow-\lambda$ due to the snap-through like instability. The point at which the disk becomes flat is associated with a minimum in the stored elastic energy, however it is not entirely dissipated. This is due to a detectable in-plane twist shear, see Fig.~\ref{fig:GA}c, inset.

We can gain further quantitative insight into the mechanism governing the transition between positive and negative Gaussian curvature through a continuum description of the steady state. In this description, the configuration of an elastic sheet of thickness $h$ is determined by its mid-plane, which is parameterized by the coordinates $u^1$ and $u^2$. Similar to the agent-based description, the active elements are considered to determine the reference state, which is encoded in the metric tensor $\overline{\bm{g}}$. We account for the bending and strain energies $E_b$ and $E_s$ of the sheet~\cite{MatozFernandez:2020cl}. Explicitly, we write
\begin{align}
\label{eq:Eb}
E_b = \frac{h^3Y}{12(1-\nu^2)}\int\textrm{d}A\left[2(H-H_0)^2 - (1-\nu)K\right]\\
\intertext{and}
\label{eq:Es}
E_s = \frac{h}{2}\int\textrm{d}A\frac{Y}{1+\nu}\left[\frac{\nu}{1-\nu}u_\alpha^\alpha u_\beta^\beta + u_\alpha^\beta u_\beta^\alpha \right].
\end{align}
Here, $Y$ is the Young's modulus and $\nu$ the Poisson ratio of the material. We set $\nu = 0.5$. The local mean and Gaussian curvatures of the surface are given by $H$ and $K$, respectively. For the time being we choose the spontaneous curvature $H_0$ to be zero, but will consider non-vanishing values below. Furthermore, $\bm{u} = (\bm{g} - \overline{\bm{g}})/2$ is the two dimensional strain tensor, where $\bm{g}$ is the 2D metric tensor. Finally, we define $u_\alpha^\beta = \overline{g}^{\beta\gamma}u_{\alpha\gamma}$. Greek indices take the values 1 and 2 and we apply Einstein's summation convention. 

In the same spirit that led to Eq.~\eqref {eq:l}, we write
\begin{multline}
\label{eq:g0}
\overline{g}_{\alpha\beta} = (1+\zeta(t))^2\big[\tilde{g}_{\alpha\beta}(1 + \xi(t) S(r))\\ + \lambda(t) S(r)[\hat{p}_\alpha \hat{p}_\beta-\frac{\tilde{g}_{\alpha\beta}}{2}]\big],
\end{multline}
where $\tilde{\bm{g}}$ is the reference metric in the absence of activity and the parameters $\zeta$, $\xi$ and $\lambda$ retain their meaning. We again take $\zeta=0$. It is worth noting that the reference metric given by Eq.~\eqref{eq:g0} may be non-embeddable. In this case the strain cannot be completely released by any sheet configuration and it remains in a state of self stress. A state of self stress is typically associated with increased stiffness and may play a role in biological systems. 

Again we study initially flat circular disks. In polar coordinates, the material's reference metric in the absence of activity is given by $\tilde{g}_{rr} = 1$,  $\tilde{g}_{\Phi\Phi} = r^2$ and $\tilde{g}_{\Phi r} = 0$. For the nematic texture given by Eq.~\eqref{eq:p} and the order parameter $S(r)=r^2/R^2$, the continuum description yields the same states as the agent-based model, see SI.

We will now consider how the reference metric changes around a topological defect with charge $q$, phase $\psi$ and anisotropic strain coefficient $\lambda$. Consider a closed loop at constant $r$. Its perimeter is
\begin{equation}
p(r) = \int_0^{2\pi}\sqrt{\overline{g}_{\Phi\Phi}}\textrm{d}\Phi
\end{equation}
and its average distance from the center 
\begin{equation}
\rho(r) = \frac{1}{2\pi}\int_0^{2\pi}\int_0^{r}\sqrt{\overline{g}_{rr}}\textrm{d}r'\textrm{d}\Phi.
\end{equation}
For a flat reference metric, $p(r) = 2\pi\rho(r)$, whereas for a surface with positive (negative) integrated Gaussian curvature $p(r) < 2\pi\rho(r)$ ($p(r) > 2\pi\rho(r)$). 

We now introduce
\begin{align}
\label{eq:kappa}
\Delta\kappa &= \frac{1}{2\pi}\left(\frac{\tilde{p}(r)}{\tilde{\rho}(r)} - \frac{p(r)}{\rho(r)}\right),
\end{align}
where $\tilde{p}(r)$ and $\tilde{\rho}(r)$ are the perimeter and radius of the loop in the absence of activity defined through $\tilde{\bm{g}}$. The quantity $\Delta\kappa$ gives us a measure of how we expect the integrated Gaussian curvature to change due to the presence of activity. 

We calculate $\Delta\kappa$ for the reference metric~\eqref{eq:g0}, nematic texture~\eqref{eq:the} with $q=1$ and order parameter $S(r)=r^2/R^2$, to obtain the values given in Fig.~\ref{fig:GA}e, inset. In particular, we find $\kappa = 0$ for $\psi = \pi/4$ for all values of $\lambda^s$. At this point the integrated azimuthal and radial strains balance and the sheet remains flat. Furthermore, $\Delta\kappa$ has an additional symmetry compared to $G$ as $\Delta\kappa(\psi,\lambda)=-\Delta\kappa(\pi/2-\psi,\lambda)$. This implies that the reduced positive Gaussian curvature observed in Fig.~\ref{fig:dyn}f and Fig.~\ref{fig:GA}d are due the sheet becoming trapped in a non-optimal configuration.

A closer look at $\Delta\kappa$ reveals why only defects with charge $q=1$ can generate Gaussian curvature. If $q\neq1$, all orientations of the director field are equally represented on the perimeter of a circle centered on the defect regardless of the phase, Eq.~\eqref{eq:the}. Thus all anisotropic expansions and contractions cancel out when integrated azimuthally, implying $\Delta\kappa=0$. In contrast, when $q=1$, then $\theta = \psi$ leading to $\Delta\kappa$ becoming a function of the phase of the defect. 

It should be noted that even for $\Delta\kappa=0$, the final reference metric $\overline{\bm g}$ may not be compatible with the starting configuration of the material $\tilde{\bm g}$. This leads to the observed increase in the stored energy also for defects with $q\neq1$, Fig.~\ref{fig:dyn}e. In particular, the reference metric can have non-zero off diagonal components leading to an in-plane shear. Specifically, when $q=1$, we have $\overline{g}_{r\Phi}=\lambda^s S(r)p_rp_\Phi = \lambda^s r^3\sin(2\psi)/(2R^2)$ leading to the chiral shear presented in Fig.~\ref{fig:GA}c, insert and SI. Thus, the chirality of the twist changes sign according to $\textrm{sign}(\psi)\times\textrm{sign}(\lambda^s)$. 

In conclusion, the phase $\psi$ can be used to control the induced Gaussian curvature and in-plane chirality when $q=1$. This is particularly important when viewed in the light of the Poincar\'{e}-Hopf theorem, which states that while the topological charge of a nematic texture is constrained by the boundaries of the material, the phase is free to vary.

\subsection{Generation of Gaussian curvature in reconstituted actomyosin gels} 
We now compare this analysis to the experimental setup of Ref.~\cite{Ideses:2018dna}. There, actin monomers, muscle myosin-II motors (MyoII), the strong cross-linker fascin, and ATP were introduced simultaneously into a sealed chamber with a lateral extension of 3.5~mm and  heights between 100~$\mu$m and 250~$\mu$m, see SI for details. Subsequently, actin polymerized and formed an elastic network with MyoII and fascin embedded. The network then contracted and quickly reached its final thickness, which could be controlled by the chamber height and the composition of the network~\cite{Ideses:2018dna}. Only afterwards, the sheets exhibited a detectable lateral contraction. After typically 5 to 10 minutes a steady state was reached.   

In the early stages of contraction, the system has rotational symmetry necessitating the existence of a single $+1$ topological defect at the center of the gel. We estimate the orientation field $\underline{\hat{p}}$ and order parameter $S$ using structure factor methods via the plugin OrientationJ for ImageJ~\cite{puspoki2016transforms} and found tangential alignment close to the gel boundary, Fig.~S11. This is consistent with an elastic active nematic disk with $q=1$, $\psi = \pi/2$. 

\begin{figure}
	\centering
	\includegraphics[width=\columnwidth]{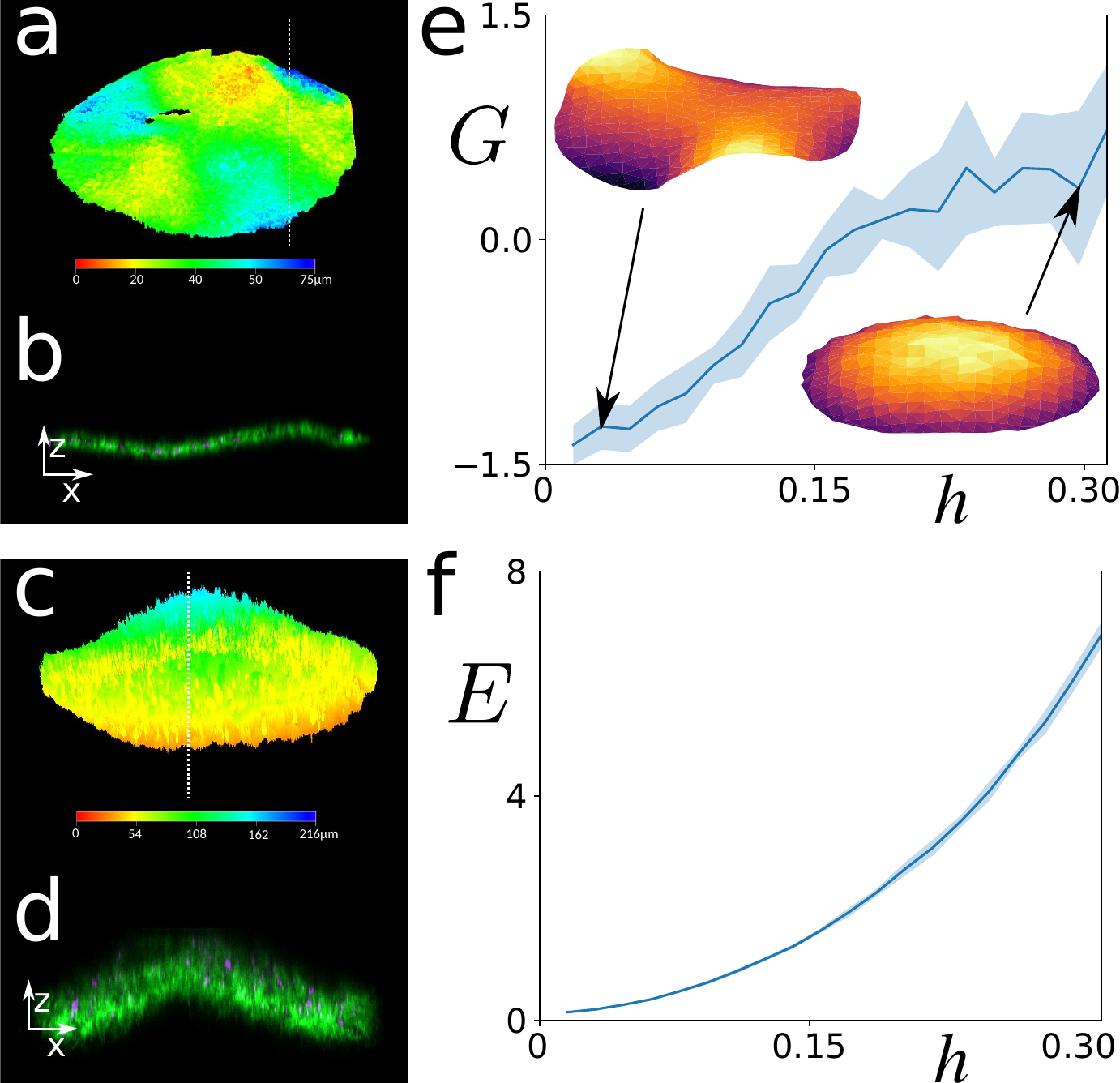}
	\caption{\label{fig:exp} Shapes of contracted actomyosin sheets containing fluorescently labeled myosin. a) Thin sheet with negative Gaussian curvature. b) Height profile of the gel (a) at the position of the dotted line. c) Thick sheet with positive Gaussian curvature. d) Height profile of the gel (c) at the position of the dotted line. e) Integrated Gaussian curvature and f) stored elastic energy of a simulated elastic active nematic material with spontaneous curvature $H_0=1$ and $\lambda^s=0.25$ as thickness is increased. }
\end{figure}

Upon contraction, thin actomyosin sheets displayed saddle configurations with multiple peaks along the periphery, Fig.~\ref{fig:exp}a,b. This would be consistent with $\lambda^s>0$, which would need to be combined with $\zeta^s<0$ to recreate the decreasing area of the gel.

Interestingly, as the thickness of the gel is increased, the number of peaks around the periphery decreases and above some critical thickness the final Gaussian curvature of the gel changes sign to be positive, Fig.~\ref{fig:exp}c,d. In the experiments, all domes buckled in the same direction, with the peak in the center pointing upwards. This suggests that the up-down symmetry of the gel was broken externally.

We use our theoretical framework to rationalize the thickness based transition from saddles to domes. First, note that the only term that breaks the up-down symmetry in the total elastic energy is the spontaneous curvature, $H_0$. We fix $H_0 = 1/R$, which favors a positive Gaussian curvature surface; a surface with negative Gaussian curvature would necessitate the divergence of the two principal curvatures and in the case of an incompressible solid, $\nu = 0.5$, increase the bending energy. 

To capture spontaneous curvature with the agent based model, we introduce a height dependent change in the rest lengths of the springs. We write
\begin{multline}
\label{eq:l2}
l_i = \tilde{l}_i(1+\zeta(t))(1+H_0z_i)\\\times\sqrt{1 +\xi(t) S(r_i) +\lambda(t) S(r_i)[(\underline{\hat{p}}\cdot \underline{\hat{l_i}})^2-0.5]},
\end{multline}
where $z_i$ is the initial elevation of the center point of spring $i$ from the mid-plane of the disk.

Consistent with our observations, we set $\psi = \pi/2$ and choose $\lambda^s > 0$. This gives $\Delta\kappa<0$ and thus induces negative Gaussian curvature. As a consequence, the bend and strain energies are frustrated with each other. These energies, Eqs.~\eqref{eq:Eb} and~\eqref{eq:Es} scale differently with the thickness $h$ of the gel, $E_b\propto h^3$ and $E_s\propto h$. Above a certain gel thickness, bending deformations are thus energetically more costly than in-plane strains. Upon simulation, we observe a similar transition between negative and positive Gaussian curvature at a critical thickness, Fig.~\ref{fig:exp}e. This transition is associated with an increase in the stored elastic energy of the gel, Fig.~\ref{fig:exp}f. We conclude that the competition between bend and strain energy can drive a transition from saddles to domes in actomyosin sheets based on the external dimensions of the gel.

The same conclusion can be reached by minimizing Eqs.~\eqref{eq:Eb} and~\eqref{eq:Es} using a Monte-Carlo method. By introducing a variational \textit{ansatz} it is also possible to predict the decreasing number of peaks around the edge of the saddle as $h$ is increased, see SI.

\subsection{Defects on elastic active nematic spheres} 
We now turn to spherical shells to demonstrate that this behavior is not limited to initially flat disks. Nematic textures on the surface of a sphere are restricted to have net topological charge $+2$ by the Poincar\'{e}-Hopf Theorem.

We utilize spherical polar coordinates with $\Theta$ being the polar angle and $\Phi$ being the azimuthal angle. These are associated with the orthonormal basis $\underline{\hat{e}}_\Theta$ and $\underline{\hat{e}}_\Phi$ of the tangent space. We again consider a thin shell with a tangential nematic texture given by 
\begin{equation}
\label{eq:p2}
\underline{\hat{p}} = {\textrm{cos}(\theta)\hat{\underline{e}}_\Theta} + {\textrm{sin}(\theta)\hat{\underline{e}}_\Phi}.
\end{equation}
We take nematic textures given by Eq.~\eqref{eq:the} with $q=1$ which results in a pair of $+1$ topological defects at opposite poles each with phase $\psi$. As before we can identify $\psi = 0$ with a pair of aster defects, $0<|\psi|<\pi/2$ with a pair of counter rotating spirals, and $\psi = \pi/2$ with a pair of vortices, see SI. We couple this with an order parameter, $S(\Theta)$, which has boundary conditions $S(0) = S(\pi) = 0$. We set it to $S(\Theta) = \cos^2(\Theta)$.

\begin{figure}[t]
	\centering
	\includegraphics[width=\columnwidth]{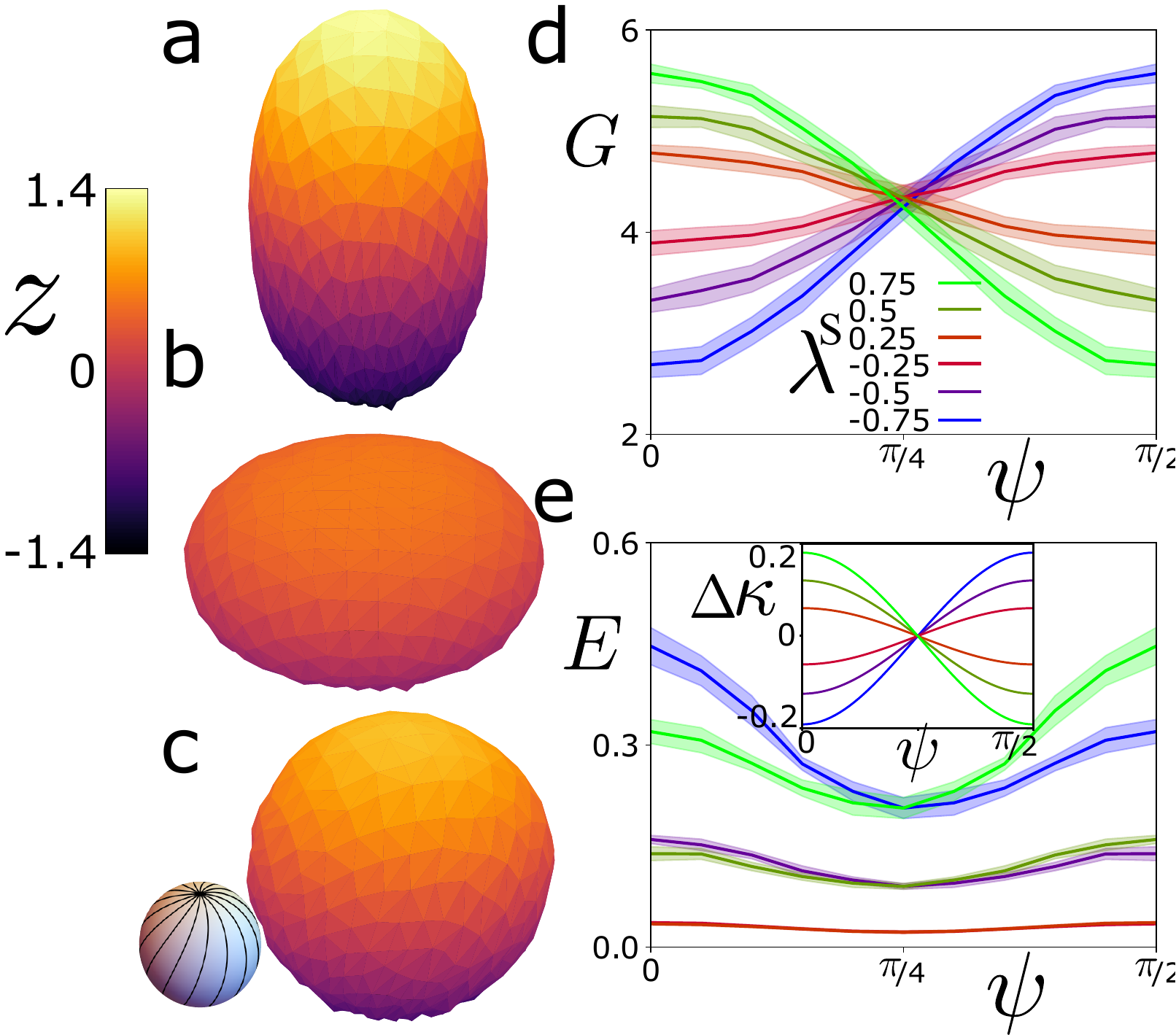}
	\caption{\label{fig:spheres} Elastic active nematic shells with spherical topology. a-c) Example surfaces for an initial spherical shell with two $+1$ defects located at opposite poles with $\psi = 0$ (a), $\psi = \pi/2$ (b), and $\psi = \pi/4$ (c). d) Integrated Gaussian curvature $G$ around the topological defects, e) stored elastic energy as a function of the phase $\psi$ for various values of $\lambda^s$. Inset: $\Delta\kappa$ for the same range of $\psi$ and $\lambda^s$.}
\end{figure}

The spherical system is evolved using the same linear growth model and Langevin dynamics previously described until a mechanical equilibrium state is reached. Here we see both oblate and prolate spheres with an intermediate twisted spherical shell, Fig.~\ref{fig:spheres}a-c. These are associated with a local decrease, increase or conservation of the Gaussian curvature at the center of each topological defect. We measure the integrated Gaussian curvature in the patches $\Theta<0.4\pi$ and $\Theta>0.6\pi$ which are centered around each topological defect. 
\begin{figure}[t]
	\centering
	\includegraphics[width=\columnwidth]{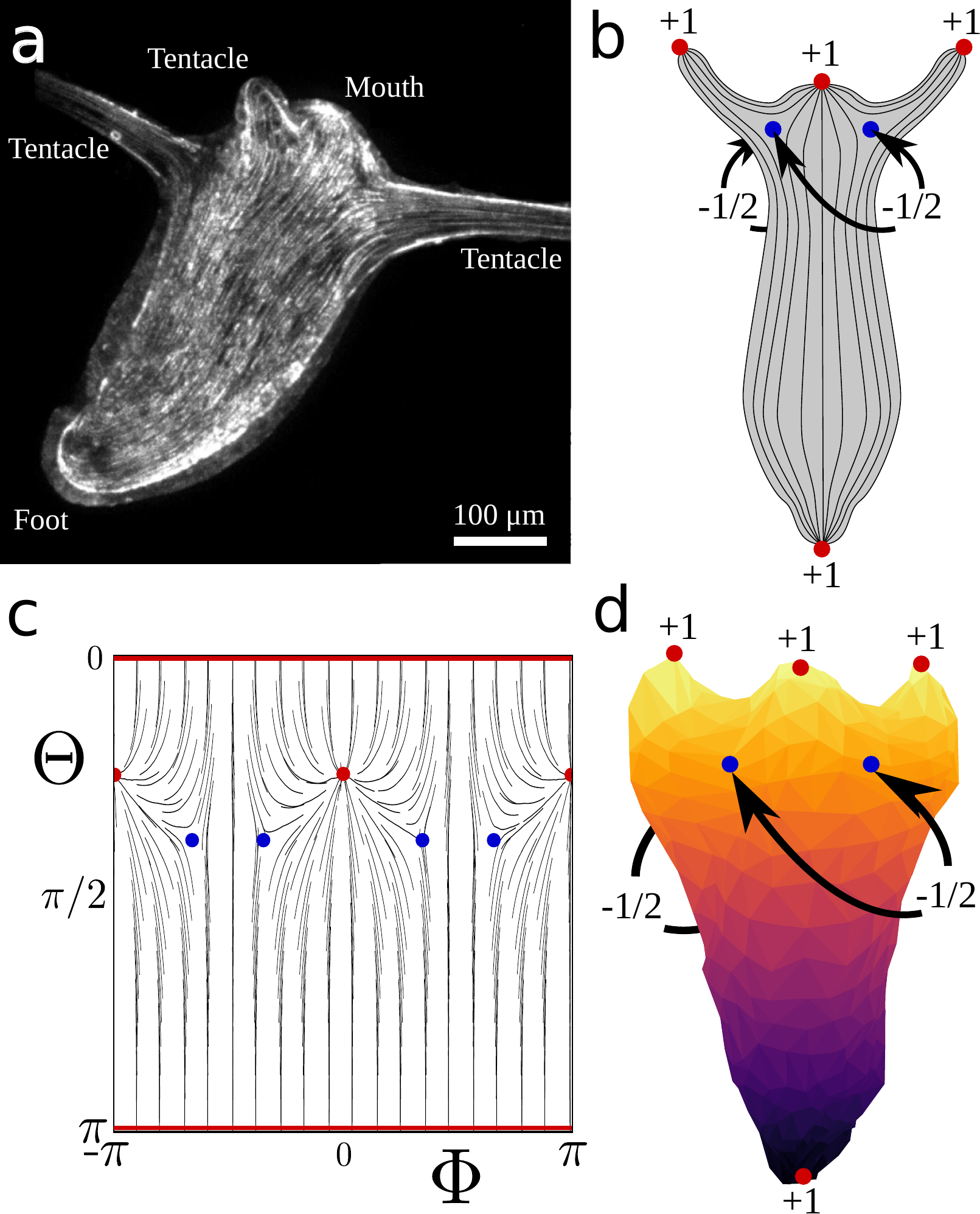}
	\caption{\label{fig:hydra} Recreating \textit{Hydra} morphology. a) Freshwater hydra with 3 tentacles. Image taken from \cite{MaroudasSacks:2020eh}. b) Schematic of the nematic texture visible in the cytoskeleton of the freshwater hydra, containing topological defects associated with key features of the animal. c) Energy minimizing nematic texture around a set of topological defects on a spherical surface. There are four $+1$ topological defects (red), the north and south pole correspond to the mouth and foot respectively. The other $+1$ defects are each associated with a pair of $-1/2$ defects (blue) which identify the future location of the tentacles. d) Final morphology of an initially spherical shell with the nematic texture given in (c). Parameter values are $\xi^s=-0.5$ and $\lambda^s=2$.}
\end{figure}

Similar to the flat disks, we see a transition between increasing and decreasing Gaussian curvature based on the sign of the active stress and the phase of the defects, Fig.~\ref{fig:spheres}d,e. Contrary to flat disks, we do not see a large asymmetry between increasing and decreasing Gaussian curvature. This is likely due to the fact that the initial Gaussian curvature on the sphere is positive, so the system does not need to spontaneously break symmetry. As in flat disks, the crossover between increasing and decreasing Gaussian curvature is associated with a local minimum in the stored elastic energy, Fig.~\ref{fig:spheres}e.

Once again this transition can be understood by calculating $\Delta\kappa$ on the surface of the sphere, Fig.~\ref{fig:spheres}e, inset. As in the case of flat disks, the transition at $\Delta\kappa=0$ is associated with an in-plane shear with a chirality depending on the sign of $\lambda^s$ and $\psi$, Fig.~\ref{fig:spheres}c, inset. 

We have shown that $+1$ topological defects are able to locally raise or lower the Gaussian curvature of the surface in which they are embedded. Additionally, we see that topological defects with $q\neq1$ do not induce a specific Gaussian curvature response. We can combine these effects to generate more complex morphologies from elastic active nematic materials. 

The organism \textit{Hydra} is a freshwater polyp with a tubular body and a number of thin tentacles around the mouth, Fig.~\ref{fig:hydra}a. \textit{Hydra}'s body wall consists of two cell layers, separated by an elastic extracellular matrix~\cite{Sarras:2012}, such that it is appropriate to describe it as an elastic active nematic  shell. The actin cytoskeleton in the outer cell layer is ordered in bundles that are largely parallel to the main axis of the animal or tentacles. This results in the existence of topological defects within the actin cytoskeleton, Fig.~\ref{fig:hydra}b. The mouth, foot and the tip of each tentacle are co-located with a $+1$ topological defect, all in the aster configuration~\cite{MaroudasSacks:2020eh}. Due to the Poincar\'{e}-Hopf theorem, the total topological charge on the animal must be $+2$, which necessitates the existence of an additional pair of $-1/2$ defects at the base of each tentacle. 

During morphogenesis, \textit{Hydra} transitions from an initially spherical shape into its final form. This involves elongation of the body between the mouth and foot and the elongation of the tentacles. This results in an increase in Gaussian curvature at each of the features associated with an aster topological defect. Considering $\Delta\kappa$, aster defects ($q=1$,$\psi=0$) are associated with an increase in Gaussian curvature during anisotropic extension ($\lambda^s > 0$). However, the additional $-1/2$ defects give $\Delta\kappa=0$ and we do not expect them to influence the local Gaussian curvature. 

We approximate the positions of the topological defects in spherical polar coordinates for a \textit{Hydra} with two tentacles in Fig.~\ref{fig:hydra}c. We reconstruct a nematic texture, $\underline{\hat{p}}$, that minimizes the elastic energy around these defects using stereographic projection from the complex plane~\cite{khoromskaia2017vortex}. We then evolve an initially spherical surface according to Eq.~\eqref{eq:l} to obtain the final shape given in Fig.~\ref{fig:hydra}d. The dynamics of this process are given in Supp. Movie 1. This figure is highly reminiscent of the final shape of \textit{Hydra} with this pattern of topological defects.

\section{Discussion}
In summary, we applied the idea that active stresses can modify the reference metric of a material in a manner linked to the orientation of the stress generating elements. We found that elastic active nematic shells featuring a $+1$ topological defect are uniquely able to induce Gaussian curvature with a sign controlled by the nature of the active stress, and the phase of defect. 

Our results have important consequences when viewed in light of the Poincar\'e-Hopf theorem. This theorem constrains the net topological charge of the defects on a surface according to its boundaries. Since $q\neq1$ defects do not change Gaussian curvature, they can be used to compensate for an arbitrary number of $q=1$ defects. These defects can thus be used freely to induce morphological changes with an arbitrary degree of complexity.

In this way, we provide further support for the idea that topological defects play a central role in guiding morphogenetic events during organismal development. In addition, these principles could be applied in the design of synthetic active materials, wherein a morphological pathway is ``programmed'' into the material by the careful arrangement of defects within the stress generating machinery~\cite{Siefert:2019bi}. Furthermore, dynamic control over the growth parameters $\zeta(t)$, $\xi(t)$ and $\lambda(t)$ could lead to materials able to undergo a sequence of morphological changes. This could be employed, for example, in the design of soft robots.

\begin{acknowledgments}
We would like to thank Nicholas Ecker for insightful discussions. GL is grateful to the Israel Ministry of Science, Technology and Space for the Jabotinsky PhD Scholarship. ABG is grateful to the Israel Science Foundation for financial support (grant  2101/20).
\end{acknowledgments}

\bibliography{PearceEtAl.bib}

\begin{thebibliography}{42}%
\makeatletter
\providecommand \@ifxundefined [1]{%
 \@ifx{#1\undefined}
}%
\providecommand \@ifnum [1]{%
 \ifnum #1\expandafter \@firstoftwo
 \else \expandafter \@secondoftwo
 \fi
}%
\providecommand \@ifx [1]{%
 \ifx #1\expandafter \@firstoftwo
 \else \expandafter \@secondoftwo
 \fi
}%
\providecommand \natexlab [1]{#1}%
\providecommand \enquote  [1]{``#1''}%
\providecommand \bibnamefont  [1]{#1}%
\providecommand \bibfnamefont [1]{#1}%
\providecommand \citenamefont [1]{#1}%
\providecommand \href@noop [0]{\@secondoftwo}%
\providecommand \href [0]{\begingroup \@sanitize@url \@href}%
\providecommand \@href[1]{\@@startlink{#1}\@@href}%
\providecommand \@@href[1]{\endgroup#1\@@endlink}%
\providecommand \@sanitize@url [0]{\catcode `\\12\catcode `\$12\catcode
  `\&12\catcode `\#12\catcode `\^12\catcode `\_12\catcode `\%12\relax}%
\providecommand \@@startlink[1]{}%
\providecommand \@@endlink[0]{}%
\providecommand \url  [0]{\begingroup\@sanitize@url \@url }%
\providecommand \@url [1]{\endgroup\@href {#1}{\urlprefix }}%
\providecommand \urlprefix  [0]{URL }%
\providecommand \Eprint [0]{\href }%
\providecommand \doibase [0]{http://dx.doi.org/}%
\providecommand \selectlanguage [0]{\@gobble}%
\providecommand \bibinfo  [0]{\@secondoftwo}%
\providecommand \bibfield  [0]{\@secondoftwo}%
\providecommand \translation [1]{[#1]}%
\providecommand \BibitemOpen [0]{}%
\providecommand \bibitemStop [0]{}%
\providecommand \bibitemNoStop [0]{.\EOS\space}%
\providecommand \EOS [0]{\spacefactor3000\relax}%
\providecommand \BibitemShut  [1]{\csname bibitem#1\endcsname}%
\let\auto@bib@innerbib\@empty
\bibitem [{\citenamefont {Alberts}\ \emph {et~al.}(2015)\citenamefont
  {Alberts}, \citenamefont {Johnson}, \citenamefont {Lewis}, \citenamefont
  {Morgan}, \citenamefont {Raff}, \citenamefont {Roberts},\ and\ \citenamefont
  {Walter}}]{Alberts:2008mo}%
  \BibitemOpen
  \bibfield  {author} {\bibinfo {author} {\bibfnamefont {B.}~\bibnamefont
  {Alberts}}, \bibinfo {author} {\bibfnamefont {A.}~\bibnamefont {Johnson}},
  \bibinfo {author} {\bibfnamefont {J.}~\bibnamefont {Lewis}}, \bibinfo
  {author} {\bibfnamefont {D.}~\bibnamefont {Morgan}}, \bibinfo {author}
  {\bibfnamefont {M.}~\bibnamefont {Raff}}, \bibinfo {author} {\bibfnamefont
  {K.}~\bibnamefont {Roberts}}, \ and\ \bibinfo {author} {\bibfnamefont
  {P.}~\bibnamefont {Walter}},\ }\href@noop {} {\emph {\bibinfo {title}
  {Molecular biology of the cell, 6th ed.}}}\ (\bibinfo  {publisher} {Garland
  Science},\ \bibinfo {address} {New York, USA},\ \bibinfo {year}
  {2015})\BibitemShut {NoStop}%
\bibitem [{\citenamefont {Gruler}\ \emph {et~al.}(1999)\citenamefont {Gruler},
  \citenamefont {Dewald},\ and\ \citenamefont {Eberhardt}}]{Gruler:1999bt}%
  \BibitemOpen
  \bibfield  {author} {\bibinfo {author} {\bibfnamefont {H.}~\bibnamefont
  {Gruler}}, \bibinfo {author} {\bibfnamefont {U.}~\bibnamefont {Dewald}}, \
  and\ \bibinfo {author} {\bibfnamefont {M.}~\bibnamefont {Eberhardt}},\ }\href
  {https://doi.org/10.1007/s100510050928} {\bibfield  {journal} {\bibinfo
  {journal} {Eur. Phys. J. B}\ }\textbf {\bibinfo {volume} {11}},\ \bibinfo
  {pages} {187} (\bibinfo {year} {1999})}\BibitemShut {NoStop}%
\bibitem [{\citenamefont {Duclos}\ \emph {et~al.}(2014)\citenamefont {Duclos},
  \citenamefont {Garcia}, \citenamefont {Yevick},\ and\ \citenamefont
  {Silberzan}}]{Duclos:2014bs}%
  \BibitemOpen
  \bibfield  {author} {\bibinfo {author} {\bibfnamefont {G.}~\bibnamefont
  {Duclos}}, \bibinfo {author} {\bibfnamefont {S.}~\bibnamefont {Garcia}},
  \bibinfo {author} {\bibfnamefont {H.~G.}\ \bibnamefont {Yevick}}, \ and\
  \bibinfo {author} {\bibfnamefont {P.}~\bibnamefont {Silberzan}},\ }\href
  {https://doi.org/10.1039/C3SM52323C} {\bibfield  {journal} {\bibinfo
  {journal} {Soft Matter}\ }\textbf {\bibinfo {volume} {10}},\ \bibinfo {pages}
  {2346} (\bibinfo {year} {2014})}\BibitemShut {NoStop}%
\bibitem [{\citenamefont {Maroudas-Sacks}\ \emph {et~al.}(2021)\citenamefont
  {Maroudas-Sacks}, \citenamefont {Garion}, \citenamefont {Shani-Zerbib},
  \citenamefont {Livshits}, \citenamefont {Braun},\ and\ \citenamefont
  {Keren}}]{MaroudasSacks:2020eh}%
  \BibitemOpen
  \bibfield  {author} {\bibinfo {author} {\bibfnamefont {Y.}~\bibnamefont
  {Maroudas-Sacks}}, \bibinfo {author} {\bibfnamefont {L.}~\bibnamefont
  {Garion}}, \bibinfo {author} {\bibfnamefont {L.}~\bibnamefont
  {Shani-Zerbib}}, \bibinfo {author} {\bibfnamefont {A.}~\bibnamefont
  {Livshits}}, \bibinfo {author} {\bibfnamefont {E.}~\bibnamefont {Braun}}, \
  and\ \bibinfo {author} {\bibfnamefont {K.}~\bibnamefont {Keren}},\ }\href
  {https://doi.org/10.1038/s41567-020-01083-1} {\bibfield  {journal} {\bibinfo
  {journal} {Nat. Phys.}\ }\textbf {\bibinfo {volume} {17}},\ \bibinfo {pages}
  {251} (\bibinfo {year} {2021})}\BibitemShut {NoStop}%
\bibitem [{\citenamefont {de~Gennes}\ and\ \citenamefont
  {Prost}(2002)}]{deGennes:2002vq}%
  \BibitemOpen
  \bibfield  {author} {\bibinfo {author} {\bibfnamefont {P.~G.}\ \bibnamefont
  {de~Gennes}}\ and\ \bibinfo {author} {\bibfnamefont {J.}~\bibnamefont
  {Prost}},\ }\href@noop {} {\emph {\bibinfo {title} {{The Physics of Liquid
  Crystals}}}},\ \bibinfo {edition} {2nd}\ ed.,\ International Series of
  Monographs on Physics\ (\bibinfo  {publisher} {Oxford University Press},\
  \bibinfo {address} {Oxford},\ \bibinfo {year} {2002})\BibitemShut {NoStop}%
\bibitem [{\citenamefont {Sanchez}\ \emph {et~al.}(2012)\citenamefont
  {Sanchez}, \citenamefont {Chen}, \citenamefont {DeCamp}, \citenamefont
  {Heymann},\ and\ \citenamefont {Dogic}}]{Sanchez:2012sp}%
  \BibitemOpen
  \bibfield  {author} {\bibinfo {author} {\bibfnamefont {T.}~\bibnamefont
  {Sanchez}}, \bibinfo {author} {\bibfnamefont {D.~T.}\ \bibnamefont {Chen}},
  \bibinfo {author} {\bibfnamefont {S.~J.}\ \bibnamefont {DeCamp}}, \bibinfo
  {author} {\bibfnamefont {M.}~\bibnamefont {Heymann}}, \ and\ \bibinfo
  {author} {\bibfnamefont {Z.}~\bibnamefont {Dogic}},\ }\href
  {https://doi.org/10.1038/nature11591} {\bibfield  {journal} {\bibinfo
  {journal} {Nature}\ }\textbf {\bibinfo {volume} {491}},\ \bibinfo {pages}
  {431} (\bibinfo {year} {2012})}\BibitemShut {NoStop}%
\bibitem [{\citenamefont {Blanch-Mercader}\ \emph
  {et~al.}(2021{\natexlab{a}})\citenamefont {Blanch-Mercader}, \citenamefont
  {Guillamat}, \citenamefont {Roux},\ and\ \citenamefont
  {Kruse}}]{BlanchMercader:2020wc}%
  \BibitemOpen
  \bibfield  {author} {\bibinfo {author} {\bibfnamefont {C.}~\bibnamefont
  {Blanch-Mercader}}, \bibinfo {author} {\bibfnamefont {P.}~\bibnamefont
  {Guillamat}}, \bibinfo {author} {\bibfnamefont {A.}~\bibnamefont {Roux}}, \
  and\ \bibinfo {author} {\bibfnamefont {K.}~\bibnamefont {Kruse}},\ }\href
  {https://doi.org/10.1103/PhysRevLett.126.028101} {\bibfield  {journal}
  {\bibinfo  {journal} {Phys. Rev. Lett.}\ }\textbf {\bibinfo {volume} {126}},\
  \bibinfo {pages} {028101} (\bibinfo {year} {2021}{\natexlab{a}})}\BibitemShut
  {NoStop}%
\bibitem [{\citenamefont {Blanch-Mercader}\ \emph
  {et~al.}(2021{\natexlab{b}})\citenamefont {Blanch-Mercader}, \citenamefont
  {Guillamat}, \citenamefont {Roux},\ and\ \citenamefont
  {Kruse}}]{BlanchMercader:2020tr}%
  \BibitemOpen
  \bibfield  {author} {\bibinfo {author} {\bibfnamefont {C.}~\bibnamefont
  {Blanch-Mercader}}, \bibinfo {author} {\bibfnamefont {P.}~\bibnamefont
  {Guillamat}}, \bibinfo {author} {\bibfnamefont {A.}~\bibnamefont {Roux}}, \
  and\ \bibinfo {author} {\bibfnamefont {K.}~\bibnamefont {Kruse}},\ }\href
  {https://doi.org/10.1103/PhysRevE.103.012405} {\bibfield  {journal} {\bibinfo
   {journal} {Phys. Rev. E}\ }\textbf {\bibinfo {volume} {103}},\ \bibinfo
  {pages} {012405} (\bibinfo {year} {2021}{\natexlab{b}})}\BibitemShut
  {NoStop}%
\bibitem [{\citenamefont {Hoffmann}\ \emph {et~al.}(2021)\citenamefont
  {Hoffmann}, \citenamefont {Carenza}, \citenamefont {Eckert},\ and\
  \citenamefont {Giomi}}]{Hoffmann:2021}%
  \BibitemOpen
  \bibfield  {author} {\bibinfo {author} {\bibfnamefont {L.~A.}\ \bibnamefont
  {Hoffmann}}, \bibinfo {author} {\bibfnamefont {L.~N.}\ \bibnamefont
  {Carenza}}, \bibinfo {author} {\bibfnamefont {J.}~\bibnamefont {Eckert}}, \
  and\ \bibinfo {author} {\bibfnamefont {L.}~\bibnamefont {Giomi}},\ }\href
  {https://arxiv.org/abs/2105.15200} {\bibfield  {journal} {\bibinfo  {journal}
  {arXiv:2105.15200v1}\ } (\bibinfo {year} {2021})}\BibitemShut {NoStop}%
\bibitem [{\citenamefont {Saw}\ \emph {et~al.}(2017)\citenamefont {Saw},
  \citenamefont {Doostmohammadi}, \citenamefont {Nier}, \citenamefont
  {Kocgozlu}, \citenamefont {Thampi}, \citenamefont {Toyama}, \citenamefont
  {Marcq}, \citenamefont {Lim}, \citenamefont {Yeomans},\ and\ \citenamefont
  {Ladoux}}]{Saw:2017gn}%
  \BibitemOpen
  \bibfield  {author} {\bibinfo {author} {\bibfnamefont {T.~B.}\ \bibnamefont
  {Saw}}, \bibinfo {author} {\bibfnamefont {A.}~\bibnamefont {Doostmohammadi}},
  \bibinfo {author} {\bibfnamefont {V.}~\bibnamefont {Nier}}, \bibinfo {author}
  {\bibfnamefont {L.}~\bibnamefont {Kocgozlu}}, \bibinfo {author}
  {\bibfnamefont {S.}~\bibnamefont {Thampi}}, \bibinfo {author} {\bibfnamefont
  {Y.}~\bibnamefont {Toyama}}, \bibinfo {author} {\bibfnamefont
  {P.}~\bibnamefont {Marcq}}, \bibinfo {author} {\bibfnamefont {C.~T.}\
  \bibnamefont {Lim}}, \bibinfo {author} {\bibfnamefont {J.~M.}\ \bibnamefont
  {Yeomans}}, \ and\ \bibinfo {author} {\bibfnamefont {B.}~\bibnamefont
  {Ladoux}},\ }\href {https://doi.org/10.1038/nature21718} {\bibfield
  {journal} {\bibinfo  {journal} {Nature}\ }\textbf {\bibinfo {volume} {544}},\
  \bibinfo {pages} {212} (\bibinfo {year} {2017})}\BibitemShut {NoStop}%
\bibitem [{\citenamefont {Kawaguchi}\ \emph {et~al.}(2017)\citenamefont
  {Kawaguchi}, \citenamefont {Kageyama},\ and\ \citenamefont
  {Sano}}]{Kawaguchi:2017em}%
  \BibitemOpen
  \bibfield  {author} {\bibinfo {author} {\bibfnamefont {K.}~\bibnamefont
  {Kawaguchi}}, \bibinfo {author} {\bibfnamefont {R.}~\bibnamefont {Kageyama}},
  \ and\ \bibinfo {author} {\bibfnamefont {M.}~\bibnamefont {Sano}},\ }\href
  {https://doi.org/10.1038/nature22321} {\bibfield  {journal} {\bibinfo
  {journal} {Nature}\ }\textbf {\bibinfo {volume} {545}},\ \bibinfo {pages}
  {327} (\bibinfo {year} {2017})}\BibitemShut {NoStop}%
\bibitem [{\citenamefont {Guillamat}\ \emph {et~al.}(2022)\citenamefont
  {Guillamat}, \citenamefont {Blanch-Mercader}, \citenamefont {Pernollet},
  \citenamefont {Kruse},\ and\ \citenamefont {Roux}}]{Guillamat:2022}%
  \BibitemOpen
  \bibfield  {author} {\bibinfo {author} {\bibfnamefont {P.}~\bibnamefont
  {Guillamat}}, \bibinfo {author} {\bibfnamefont {C.}~\bibnamefont
  {Blanch-Mercader}}, \bibinfo {author} {\bibfnamefont {G.}~\bibnamefont
  {Pernollet}}, \bibinfo {author} {\bibfnamefont {K.}~\bibnamefont {Kruse}}, \
  and\ \bibinfo {author} {\bibfnamefont {A.}~\bibnamefont {Roux}},\ }\href
  {\doibase 10.1038/s41563-022-01194-5} {\bibfield  {journal} {\bibinfo
  {journal} {Nature Materials}\ }\textbf {\bibinfo {volume} {21}},\ \bibinfo
  {pages} {588} (\bibinfo {year} {2022})}\BibitemShut {NoStop}%
\bibitem [{\citenamefont {Liu}\ and\ \citenamefont
  {Fletcher}(2009)}]{Liu:2009da}%
  \BibitemOpen
  \bibfield  {author} {\bibinfo {author} {\bibfnamefont {A.~P.}\ \bibnamefont
  {Liu}}\ and\ \bibinfo {author} {\bibfnamefont {D.~A.}\ \bibnamefont
  {Fletcher}},\ }\href {https://doi.org/10.1038/nrm2746} {\bibfield  {journal}
  {\bibinfo  {journal} {Nat. Rev. Mol. Cell Bio.}\ }\textbf {\bibinfo {volume}
  {10}},\ \bibinfo {pages} {644} (\bibinfo {year} {2009})}\BibitemShut
  {NoStop}%
\bibitem [{\citenamefont {Cameron}\ \emph {et~al.}(1999)\citenamefont
  {Cameron}, \citenamefont {Footer}, \citenamefont {van Oudenaarden},\ and\
  \citenamefont {Theriot}}]{Cameron:1999}%
  \BibitemOpen
  \bibfield  {author} {\bibinfo {author} {\bibfnamefont {L.~A.}\ \bibnamefont
  {Cameron}}, \bibinfo {author} {\bibfnamefont {M.~J.}\ \bibnamefont {Footer}},
  \bibinfo {author} {\bibfnamefont {A.}~\bibnamefont {van Oudenaarden}}, \ and\
  \bibinfo {author} {\bibfnamefont {J.~A.}\ \bibnamefont {Theriot}},\ }\href
  {https://doi.org/10.1073/pnas.96.9.4908} {\bibfield  {journal} {\bibinfo
  {journal} {Proc. Natl. Acad. Sci. USA}\ }\textbf {\bibinfo {volume} {96}},\
  \bibinfo {pages} {4908} (\bibinfo {year} {1999})}\BibitemShut {NoStop}%
\bibitem [{\citenamefont {Bernheim-Groswasser}\ \emph
  {et~al.}(2002)\citenamefont {Bernheim-Groswasser}, \citenamefont {Wiesner},
  \citenamefont {Golsteyn}, \citenamefont {Carlier},\ and\ \citenamefont
  {Sykes}}]{Bernheim:2002}%
  \BibitemOpen
  \bibfield  {author} {\bibinfo {author} {\bibfnamefont {A.}~\bibnamefont
  {Bernheim-Groswasser}}, \bibinfo {author} {\bibfnamefont {S.}~\bibnamefont
  {Wiesner}}, \bibinfo {author} {\bibfnamefont {R.~M.}\ \bibnamefont
  {Golsteyn}}, \bibinfo {author} {\bibfnamefont {M.~F.}\ \bibnamefont
  {Carlier}}, \ and\ \bibinfo {author} {\bibfnamefont {C.}~\bibnamefont
  {Sykes}},\ }\href {https://doi.org/10.1038/417308a} {\bibfield  {journal}
  {\bibinfo  {journal} {Nature}\ }\textbf {\bibinfo {volume} {417}},\ \bibinfo
  {pages} {308} (\bibinfo {year} {2002})}\BibitemShut {NoStop}%
\bibitem [{\citenamefont {Dayel}\ \emph {et~al.}(2009)\citenamefont {Dayel},
  \citenamefont {Akin}, \citenamefont {Landeryou}, \citenamefont {Risca},
  \citenamefont {Mogilner},\ and\ \citenamefont {Mullins}}]{Dayel:2009}%
  \BibitemOpen
  \bibfield  {author} {\bibinfo {author} {\bibfnamefont {M.~J.}\ \bibnamefont
  {Dayel}}, \bibinfo {author} {\bibfnamefont {O.}~\bibnamefont {Akin}},
  \bibinfo {author} {\bibfnamefont {M.}~\bibnamefont {Landeryou}}, \bibinfo
  {author} {\bibfnamefont {V.}~\bibnamefont {Risca}}, \bibinfo {author}
  {\bibfnamefont {A.}~\bibnamefont {Mogilner}}, \ and\ \bibinfo {author}
  {\bibfnamefont {R.~D.}\ \bibnamefont {Mullins}},\ }\href
  {https://doi.org/10.1371/journal.pbio.1000201} {\bibfield  {journal}
  {\bibinfo  {journal} {PLoS Biol.}\ }\textbf {\bibinfo {volume} {7}},\
  \bibinfo {pages} {e1000201} (\bibinfo {year} {2009})}\BibitemShut {NoStop}%
\bibitem [{\citenamefont {Siton}\ \emph {et~al.}(2011)\citenamefont {Siton},
  \citenamefont {Ideses}, \citenamefont {Albeck}, \citenamefont {Unger},
  \citenamefont {Bershadsky}, \citenamefont {Gov},\ and\ \citenamefont
  {Bernheim-Groswasser}}]{Siton:2011}%
  \BibitemOpen
  \bibfield  {author} {\bibinfo {author} {\bibfnamefont {O.}~\bibnamefont
  {Siton}}, \bibinfo {author} {\bibfnamefont {Y.}~\bibnamefont {Ideses}},
  \bibinfo {author} {\bibfnamefont {S.}~\bibnamefont {Albeck}}, \bibinfo
  {author} {\bibfnamefont {T.}~\bibnamefont {Unger}}, \bibinfo {author}
  {\bibfnamefont {A.~D.}\ \bibnamefont {Bershadsky}}, \bibinfo {author}
  {\bibfnamefont {N.~S.}\ \bibnamefont {Gov}}, \ and\ \bibinfo {author}
  {\bibfnamefont {A.}~\bibnamefont {Bernheim-Groswasser}},\ }\href
  {https://doi.org/10.1016/j.cub.2011.11.010} {\bibfield  {journal} {\bibinfo
  {journal} {Curr. Biol.}\ }\textbf {\bibinfo {volume} {21}},\ \bibinfo {pages}
  {2092} (\bibinfo {year} {2011})}\BibitemShut {NoStop}%
\bibitem [{\citenamefont {Backouche}\ \emph {et~al.}(2006)\citenamefont
  {Backouche}, \citenamefont {Haviv}, \citenamefont {Groswasser},\ and\
  \citenamefont {Bernheim-Groswasser}}]{Backouche:2006}%
  \BibitemOpen
  \bibfield  {author} {\bibinfo {author} {\bibfnamefont {F.}~\bibnamefont
  {Backouche}}, \bibinfo {author} {\bibfnamefont {L.}~\bibnamefont {Haviv}},
  \bibinfo {author} {\bibfnamefont {D.}~\bibnamefont {Groswasser}}, \ and\
  \bibinfo {author} {\bibfnamefont {A.}~\bibnamefont {Bernheim-Groswasser}},\
  }\href {https://doi.org/10.1088/1478-3975/3/4/004} {\bibfield  {journal}
  {\bibinfo  {journal} {Phys. Biol.}\ }\textbf {\bibinfo {volume} {3}},\
  \bibinfo {pages} {264} (\bibinfo {year} {2006})}\BibitemShut {NoStop}%
\bibitem [{\citenamefont {Reymann}\ \emph {et~al.}(2012)\citenamefont
  {Reymann}, \citenamefont {Boujemaa-Paterski}, \citenamefont {Martiel},
  \citenamefont {Gu\'erin}, \citenamefont {Cao}, \citenamefont {Chin},
  \citenamefont {De~La~Cruz}, \citenamefont {Th\'ery},\ and\ \citenamefont
  {Blanchoin}}]{Reymann:2012}%
  \BibitemOpen
  \bibfield  {author} {\bibinfo {author} {\bibfnamefont {A.-C.}\ \bibnamefont
  {Reymann}}, \bibinfo {author} {\bibfnamefont {R.}~\bibnamefont
  {Boujemaa-Paterski}}, \bibinfo {author} {\bibfnamefont {J.-L.}\ \bibnamefont
  {Martiel}}, \bibinfo {author} {\bibfnamefont {C.}~\bibnamefont {Gu\'erin}},
  \bibinfo {author} {\bibfnamefont {W.}~\bibnamefont {Cao}}, \bibinfo {author}
  {\bibfnamefont {H.~F.}\ \bibnamefont {Chin}}, \bibinfo {author}
  {\bibfnamefont {E.~M.}\ \bibnamefont {De~La~Cruz}}, \bibinfo {author}
  {\bibfnamefont {M.}~\bibnamefont {Th\'ery}}, \ and\ \bibinfo {author}
  {\bibfnamefont {L.}~\bibnamefont {Blanchoin}},\ }\href
  {https://doi.org/10.1126/science.1221708} {\bibfield  {journal} {\bibinfo
  {journal} {Science}\ }\textbf {\bibinfo {volume} {336}},\ \bibinfo {pages}
  {1310} (\bibinfo {year} {2012})}\BibitemShut {NoStop}%
\bibitem [{\citenamefont {K\"ohler}\ and\ \citenamefont
  {Bausch}(2012)}]{Kohler:2012}%
  \BibitemOpen
  \bibfield  {author} {\bibinfo {author} {\bibfnamefont {S.}~\bibnamefont
  {K\"ohler}}\ and\ \bibinfo {author} {\bibfnamefont {A.~R.}\ \bibnamefont
  {Bausch}},\ }\href {https://doi.org/10.1371/journal.pone.0039869} {\bibfield
  {journal} {\bibinfo  {journal} {PLoS ONE}\ }\textbf {\bibinfo {volume} {7}},\
  \bibinfo {pages} {e39869} (\bibinfo {year} {2012})}\BibitemShut {NoStop}%
\bibitem [{\citenamefont {Ideses}\ \emph {et~al.}(2013)\citenamefont {Ideses},
  \citenamefont {Sonn-Segev}, \citenamefont {Roichman},\ and\ \citenamefont
  {Bernheim-Groswasser}}]{Ideses:2013fr}%
  \BibitemOpen
  \bibfield  {author} {\bibinfo {author} {\bibfnamefont {Y.}~\bibnamefont
  {Ideses}}, \bibinfo {author} {\bibfnamefont {A.}~\bibnamefont {Sonn-Segev}},
  \bibinfo {author} {\bibfnamefont {Y.}~\bibnamefont {Roichman}}, \ and\
  \bibinfo {author} {\bibfnamefont {A.}~\bibnamefont {Bernheim-Groswasser}},\
  }\href {https://doi.org/10.1039/C3SM50309G} {\bibfield  {journal} {\bibinfo
  {journal} {Soft Matter}\ }\textbf {\bibinfo {volume} {9}},\ \bibinfo {pages}
  {7127} (\bibinfo {year} {2013})}\BibitemShut {NoStop}%
\bibitem [{\citenamefont {Alvarado}\ \emph {et~al.}(2013)\citenamefont
  {Alvarado}, \citenamefont {Sheinman}, \citenamefont {Sharma}, \citenamefont
  {MacKintosh},\ and\ \citenamefont {Koenderink}}]{Alvarado:2013}%
  \BibitemOpen
  \bibfield  {author} {\bibinfo {author} {\bibfnamefont {J.}~\bibnamefont
  {Alvarado}}, \bibinfo {author} {\bibfnamefont {M.}~\bibnamefont {Sheinman}},
  \bibinfo {author} {\bibfnamefont {A.}~\bibnamefont {Sharma}}, \bibinfo
  {author} {\bibfnamefont {F.~C.}\ \bibnamefont {MacKintosh}}, \ and\ \bibinfo
  {author} {\bibfnamefont {G.~H.}\ \bibnamefont {Koenderink}},\ }\href
  {https://doi.org/10.1038/nphys2715} {\bibfield  {journal} {\bibinfo
  {journal} {Nat. Phys.}\ }\textbf {\bibinfo {volume} {9}},\ \bibinfo {pages}
  {591} (\bibinfo {year} {2013})}\BibitemShut {NoStop}%
\bibitem [{\citenamefont {Linsmeier}\ \emph {et~al.}(2016)\citenamefont
  {Linsmeier}, \citenamefont {Banerjee}, \citenamefont {Oakes}, \citenamefont
  {Jung}, \citenamefont {Kim},\ and\ \citenamefont {Murrell}}]{Linsmeier:2016}%
  \BibitemOpen
  \bibfield  {author} {\bibinfo {author} {\bibfnamefont {I.}~\bibnamefont
  {Linsmeier}}, \bibinfo {author} {\bibfnamefont {S.}~\bibnamefont {Banerjee}},
  \bibinfo {author} {\bibfnamefont {P.~W.}\ \bibnamefont {Oakes}}, \bibinfo
  {author} {\bibfnamefont {W.}~\bibnamefont {Jung}}, \bibinfo {author}
  {\bibfnamefont {T.}~\bibnamefont {Kim}}, \ and\ \bibinfo {author}
  {\bibfnamefont {M.~P.}\ \bibnamefont {Murrell}},\ }\href
  {https://doi.org/10.1038/ncomms12615} {\bibfield  {journal} {\bibinfo
  {journal} {Nat. Commun.}\ }\textbf {\bibinfo {volume} {7}},\ \bibinfo {pages}
  {12615} (\bibinfo {year} {2016})}\BibitemShut {NoStop}%
\bibitem [{\citenamefont {Ennomani}\ \emph {et~al.}(2016)\citenamefont
  {Ennomani}, \citenamefont {Letort}, \citenamefont {Gu\'erin}, \citenamefont
  {Martiel}, \citenamefont {Cao}, \citenamefont {N\'ed\'elec}, \citenamefont
  {Enrique}, \citenamefont {Th\'ery},\ and\ \citenamefont
  {Blanchoin}}]{Ennomani:2016}%
  \BibitemOpen
  \bibfield  {author} {\bibinfo {author} {\bibfnamefont {H.}~\bibnamefont
  {Ennomani}}, \bibinfo {author} {\bibfnamefont {G.}~\bibnamefont {Letort}},
  \bibinfo {author} {\bibfnamefont {C.}~\bibnamefont {Gu\'erin}}, \bibinfo
  {author} {\bibfnamefont {J.-L.}\ \bibnamefont {Martiel}}, \bibinfo {author}
  {\bibfnamefont {W.}~\bibnamefont {Cao}}, \bibinfo {author} {\bibfnamefont
  {F.}~\bibnamefont {N\'ed\'elec}}, \bibinfo {author} {\bibfnamefont
  {M.}~\bibnamefont {Enrique}}, \bibinfo {author} {\bibfnamefont
  {M.}~\bibnamefont {Th\'ery}}, \ and\ \bibinfo {author} {\bibfnamefont
  {L.}~\bibnamefont {Blanchoin}},\ }\href
  {https://doi.org/10.1016/j.cub.2015.12.069} {\bibfield  {journal} {\bibinfo
  {journal} {Curr. Biol.}\ }\textbf {\bibinfo {volume} {26}},\ \bibinfo {pages}
  {616} (\bibinfo {year} {2016})}\BibitemShut {NoStop}%
\bibitem [{\citenamefont {Schuppler}\ \emph {et~al.}(2016)\citenamefont
  {Schuppler}, \citenamefont {Keber}, \citenamefont {Kr\"oger},\ and\
  \citenamefont {Bausch}}]{Schuppler:2016}%
  \BibitemOpen
  \bibfield  {author} {\bibinfo {author} {\bibfnamefont {M.}~\bibnamefont
  {Schuppler}}, \bibinfo {author} {\bibfnamefont {F.~C.}\ \bibnamefont
  {Keber}}, \bibinfo {author} {\bibfnamefont {M.}~\bibnamefont {Kr\"oger}}, \
  and\ \bibinfo {author} {\bibfnamefont {A.~R.}\ \bibnamefont {Bausch}},\
  }\href {https://doi.org/10.1038/ncomms13120} {\bibfield  {journal} {\bibinfo
  {journal} {Nat. Commun.}\ }\textbf {\bibinfo {volume} {7}},\ \bibinfo {pages}
  {13120} (\bibinfo {year} {2016})}\BibitemShut {NoStop}%
\bibitem [{\citenamefont {Boukellal}\ \emph {et~al.}(2004)\citenamefont
  {Boukellal}, \citenamefont {Campas}, \citenamefont {Joanny}, \citenamefont
  {Prost},\ and\ \citenamefont {Sykes}}]{Boukellal:2004}%
  \BibitemOpen
  \bibfield  {author} {\bibinfo {author} {\bibfnamefont {H.}~\bibnamefont
  {Boukellal}}, \bibinfo {author} {\bibfnamefont {O.}~\bibnamefont {Campas}},
  \bibinfo {author} {\bibfnamefont {J.~F.}\ \bibnamefont {Joanny}}, \bibinfo
  {author} {\bibfnamefont {J.}~\bibnamefont {Prost}}, \ and\ \bibinfo {author}
  {\bibfnamefont {C.}~\bibnamefont {Sykes}},\ }\href
  {https://doi.org/10.1103/PhysRevE.69.061906} {\bibfield  {journal} {\bibinfo
  {journal} {Phys. Rev. E}\ }\textbf {\bibinfo {volume} {69}},\ \bibinfo
  {pages} {061906} (\bibinfo {year} {2004})}\BibitemShut {NoStop}%
\bibitem [{\citenamefont {Ideses}\ \emph {et~al.}(2018)\citenamefont {Ideses},
  \citenamefont {Erukhimovitch}, \citenamefont {Brand}, \citenamefont
  {Jourdain}, \citenamefont {Hernandez}, \citenamefont {Gabinet}, \citenamefont
  {Safran}, \citenamefont {Kruse},\ and\ \citenamefont
  {Bernheim-Groswasser}}]{Ideses:2018dna}%
  \BibitemOpen
  \bibfield  {author} {\bibinfo {author} {\bibfnamefont {Y.}~\bibnamefont
  {Ideses}}, \bibinfo {author} {\bibfnamefont {V.}~\bibnamefont
  {Erukhimovitch}}, \bibinfo {author} {\bibfnamefont {R.}~\bibnamefont
  {Brand}}, \bibinfo {author} {\bibfnamefont {D.}~\bibnamefont {Jourdain}},
  \bibinfo {author} {\bibfnamefont {J.~S.}\ \bibnamefont {Hernandez}}, \bibinfo
  {author} {\bibfnamefont {U.~R.}\ \bibnamefont {Gabinet}}, \bibinfo {author}
  {\bibfnamefont {S.~A.}\ \bibnamefont {Safran}}, \bibinfo {author}
  {\bibfnamefont {K.}~\bibnamefont {Kruse}}, \ and\ \bibinfo {author}
  {\bibfnamefont {A.}~\bibnamefont {Bernheim-Groswasser}},\ }\href
  {https://doi.org/10.1038/s41467-018-04829-x} {\bibfield  {journal} {\bibinfo
  {journal} {Nat. Commun.}\ }\textbf {\bibinfo {volume} {9}},\ \bibinfo {pages}
  {2461} (\bibinfo {year} {2018})}\BibitemShut {NoStop}%
\bibitem [{\citenamefont {Zakharov}\ and\ \citenamefont
  {Dasbiswas}(2021{\natexlab{a}})}]{Zakharov:2021a}%
  \BibitemOpen
  \bibfield  {author} {\bibinfo {author} {\bibfnamefont {A.}~\bibnamefont
  {Zakharov}}\ and\ \bibinfo {author} {\bibfnamefont {K.}~\bibnamefont
  {Dasbiswas}},\ }\href {\doibase 10.1140/epje/s10189-021-00086-x} {\bibfield
  {journal} {\bibinfo  {journal} {Eur. Phys. J. E}\ }\textbf {\bibinfo {volume}
  {44}},\ \bibinfo {pages} {82} (\bibinfo {year}
  {2021}{\natexlab{a}})}\BibitemShut {NoStop}%
\bibitem [{\citenamefont {Zakharov}\ and\ \citenamefont
  {Dasbiswas}(2021{\natexlab{b}})}]{Zakharov:2021b}%
  \BibitemOpen
  \bibfield  {author} {\bibinfo {author} {\bibfnamefont {A.}~\bibnamefont
  {Zakharov}}\ and\ \bibinfo {author} {\bibfnamefont {K.}~\bibnamefont
  {Dasbiswas}},\ }\href {\doibase 10.1039/D1SM00003A} {\bibfield  {journal}
  {\bibinfo  {journal} {Soft Matter}\ }\textbf {\bibinfo {volume} {17}},\
  \bibinfo {pages} {4738} (\bibinfo {year} {2021}{\natexlab{b}})}\BibitemShut
  {NoStop}%
\bibitem [{\citenamefont {Matoz-Fernandez}\ \emph {et~al.}(2020)\citenamefont
  {Matoz-Fernandez}, \citenamefont {Davidson}, \citenamefont {Stanley-Wall},\
  and\ \citenamefont {Sknepnek}}]{MatozFernandez:2020cl}%
  \BibitemOpen
  \bibfield  {author} {\bibinfo {author} {\bibfnamefont {D.~A.}\ \bibnamefont
  {Matoz-Fernandez}}, \bibinfo {author} {\bibfnamefont {F.~A.}\ \bibnamefont
  {Davidson}}, \bibinfo {author} {\bibfnamefont {N.~R.}\ \bibnamefont
  {Stanley-Wall}}, \ and\ \bibinfo {author} {\bibfnamefont {R.}~\bibnamefont
  {Sknepnek}},\ }\href {https://doi.org/10.1103/PhysRevResearch.2.013165}
  {\bibfield  {journal} {\bibinfo  {journal} {Phys. Rev. Research}\ }\textbf
  {\bibinfo {volume} {2}},\ \bibinfo {pages} {013165} (\bibinfo {year}
  {2020})}\BibitemShut {NoStop}%
\bibitem [{\citenamefont {Berthoumieux}\ \emph {et~al.}(2014)\citenamefont
  {Berthoumieux}, \citenamefont {Maitre}, \citenamefont {Heisenberg},
  \citenamefont {Paluch}, \citenamefont {J\"ulicher},\ and\ \citenamefont
  {Salbreux}}]{Berthoumieux:2014eo}%
  \BibitemOpen
  \bibfield  {author} {\bibinfo {author} {\bibfnamefont {H.}~\bibnamefont
  {Berthoumieux}}, \bibinfo {author} {\bibfnamefont {J.-L.}\ \bibnamefont
  {Maitre}}, \bibinfo {author} {\bibfnamefont {C.-P.}\ \bibnamefont
  {Heisenberg}}, \bibinfo {author} {\bibfnamefont {E.~K.}\ \bibnamefont
  {Paluch}}, \bibinfo {author} {\bibfnamefont {F.}~\bibnamefont {J\"ulicher}},
  \ and\ \bibinfo {author} {\bibfnamefont {G.}~\bibnamefont {Salbreux}},\
  }\href {https://doi.org/10.1088/1367-2630/16/6/065005} {\bibfield  {journal}
  {\bibinfo  {journal} {New J. Phys.}\ }\textbf {\bibinfo {volume} {16}},\
  \bibinfo {pages} {065005} (\bibinfo {year} {2014})}\BibitemShut {NoStop}%
\bibitem [{\citenamefont {Salbreux}\ and\ \citenamefont
  {J\"ulicher}(2017)}]{Salbreux:2017ci}%
  \BibitemOpen
  \bibfield  {author} {\bibinfo {author} {\bibfnamefont {G.}~\bibnamefont
  {Salbreux}}\ and\ \bibinfo {author} {\bibfnamefont {F.}~\bibnamefont
  {J\"ulicher}},\ }\href {https://doi.org/10.1103/PhysRevE.96.032404}
  {\bibfield  {journal} {\bibinfo  {journal} {Phys. Rev. E}\ }\textbf {\bibinfo
  {volume} {96}},\ \bibinfo {pages} {032404} (\bibinfo {year}
  {2017})}\BibitemShut {NoStop}%
\bibitem [{\citenamefont {Morris}\ and\ \citenamefont
  {Rao}(2019)}]{Morris:2019bx}%
  \BibitemOpen
  \bibfield  {author} {\bibinfo {author} {\bibfnamefont {R.~G.}\ \bibnamefont
  {Morris}}\ and\ \bibinfo {author} {\bibfnamefont {M.}~\bibnamefont {Rao}},\
  }\href {https://doi.org/10.1103/PhysRevE.100.022413} {\bibfield  {journal}
  {\bibinfo  {journal} {Phys. Rev. E}\ }\textbf {\bibinfo {volume} {100}},\
  \bibinfo {pages} {022413} (\bibinfo {year} {2019})}\BibitemShut {NoStop}%
\bibitem [{\citenamefont {Frank}\ and\ \citenamefont
  {Kardar}(2008)}]{Frank:2008}%
  \BibitemOpen
  \bibfield  {author} {\bibinfo {author} {\bibfnamefont {J.~R.}\ \bibnamefont
  {Frank}}\ and\ \bibinfo {author} {\bibfnamefont {M.}~\bibnamefont {Kardar}},\
  }\href {https://doi.org/10.1103/PhysRevE.77.041705} {\bibfield  {journal}
  {\bibinfo  {journal} {Phys. Rev. E}\ }\textbf {\bibinfo {volume} {77}},\
  \bibinfo {pages} {041705} (\bibinfo {year} {2008})}\BibitemShut {NoStop}%
\bibitem [{\citenamefont {Modes}\ and\ \citenamefont
  {Warner}(2011)}]{Modes:2011}%
  \BibitemOpen
  \bibfield  {author} {\bibinfo {author} {\bibfnamefont {C.~D.}\ \bibnamefont
  {Modes}}\ and\ \bibinfo {author} {\bibfnamefont {M.}~\bibnamefont {Warner}},\
  }\href {https://doi.org/10.1103/PhysRevE.84.021711} {\bibfield  {journal}
  {\bibinfo  {journal} {Phys. Rev. E}\ }\textbf {\bibinfo {volume} {84}},\
  \bibinfo {pages} {021711} (\bibinfo {year} {2011})}\BibitemShut {NoStop}%
\bibitem [{\citenamefont {Zhang}\ \emph {et~al.}(2018)\citenamefont {Zhang},
  \citenamefont {Ross}, \citenamefont {Gardel},\ and\ \citenamefont
  {de~Pablo}}]{Zhang:2018}%
  \BibitemOpen
  \bibfield  {author} {\bibinfo {author} {\bibfnamefont {R.}~\bibnamefont
  {Zhang}}, \bibinfo {author} {\bibfnamefont {J.~L.}\ \bibnamefont {Ross}},
  \bibinfo {author} {\bibfnamefont {M.~L.}\ \bibnamefont {Gardel}}, \ and\
  \bibinfo {author} {\bibfnamefont {J.~J.}\ \bibnamefont {de~Pablo}},\ }\href
  {https://doi.org/10.1073/pnas.1713832115} {\bibfield  {journal} {\bibinfo
  {journal} {Proc. Natl. Acad. Sci. USA}\ }\textbf {\bibinfo {volume} {115}},\
  \bibinfo {pages} {E124} (\bibinfo {year} {2018})}\BibitemShut {NoStop}%
\bibitem [{\citenamefont {Pearce}\ \emph {et~al.}(2019)\citenamefont {Pearce},
  \citenamefont {Ellis}, \citenamefont {Fernandez-Nieves},\ and\ \citenamefont
  {Giomi}}]{pearce2019geometrical}%
  \BibitemOpen
  \bibfield  {author} {\bibinfo {author} {\bibfnamefont {D.~J.~G.}\
  \bibnamefont {Pearce}}, \bibinfo {author} {\bibfnamefont {P.~W.}\
  \bibnamefont {Ellis}}, \bibinfo {author} {\bibfnamefont {A.}~\bibnamefont
  {Fernandez-Nieves}}, \ and\ \bibinfo {author} {\bibfnamefont
  {L.}~\bibnamefont {Giomi}},\ }\href {\doibase 10.1103/physrevlett.122.168002}
  {\bibfield  {journal} {\bibinfo  {journal} {Phys. Rev. Lett.}\ }\textbf
  {\bibinfo {volume} {122}},\ \bibinfo {pages} {168002} (\bibinfo {year}
  {2019})},\ \Eprint {http://arxiv.org/abs/1805.01455} {1805.01455}
  \BibitemShut {NoStop}%
\bibitem [{\citenamefont {Ellis}\ \emph {et~al.}(2018)\citenamefont {Ellis},
  \citenamefont {Pearce}, \citenamefont {Chang}, \citenamefont {Goldsztein},
  \citenamefont {Giomi},\ and\ \citenamefont
  {Fernandez-Nieves}}]{ellis2018curvature}%
  \BibitemOpen
  \bibfield  {author} {\bibinfo {author} {\bibfnamefont {P.~W.}\ \bibnamefont
  {Ellis}}, \bibinfo {author} {\bibfnamefont {D.~J.~G.}\ \bibnamefont
  {Pearce}}, \bibinfo {author} {\bibfnamefont {Y.-W.}\ \bibnamefont {Chang}},
  \bibinfo {author} {\bibfnamefont {G.}~\bibnamefont {Goldsztein}}, \bibinfo
  {author} {\bibfnamefont {L.}~\bibnamefont {Giomi}}, \ and\ \bibinfo {author}
  {\bibfnamefont {A.}~\bibnamefont {Fernandez-Nieves}},\ }\href {\doibase
  10.1038/nphys4276} {\bibfield  {journal} {\bibinfo  {journal} {Nat. Phys.}\
  }\textbf {\bibinfo {volume} {14}},\ \bibinfo {pages} {85 } (\bibinfo {year}
  {2018})}\BibitemShut {NoStop}%
\bibitem [{\citenamefont {P{\"u}sp{\"o}ki}\ \emph {et~al.}(2016)\citenamefont
  {P{\"u}sp{\"o}ki}, \citenamefont {Storath}, \citenamefont {Sage},\ and\
  \citenamefont {Unser}}]{puspoki2016transforms}%
  \BibitemOpen
  \bibfield  {author} {\bibinfo {author} {\bibfnamefont {Z.}~\bibnamefont
  {P{\"u}sp{\"o}ki}}, \bibinfo {author} {\bibfnamefont {M.}~\bibnamefont
  {Storath}}, \bibinfo {author} {\bibfnamefont {D.}~\bibnamefont {Sage}}, \
  and\ \bibinfo {author} {\bibfnamefont {M.}~\bibnamefont {Unser}},\ }\href
  {\doibase 10.1007/978-3-319-28549-8\_3} {\bibfield  {journal} {\bibinfo
  {journal} {Advances in Anatomy, Embryology and Cell Biology}\ }\textbf
  {\bibinfo {volume} {219}},\ \bibinfo {pages} {69} (\bibinfo {year}
  {2016})}\BibitemShut {NoStop}%
\bibitem [{\citenamefont {Sarras}(2012)}]{Sarras:2012}%
  \BibitemOpen
  \bibfield  {author} {\bibinfo {author} {\bibfnamefont {M.~P.}\ \bibnamefont
  {Sarras}},\ }\href {\doibase 10.1387/ijdb.113445ms} {\bibfield  {journal}
  {\bibinfo  {journal} {Int. J. Dev. Biol.}\ }\textbf {\bibinfo {volume}
  {56}},\ \bibinfo {pages} {567} (\bibinfo {year} {2012})}\BibitemShut
  {NoStop}%
\bibitem [{\citenamefont {Khoromskaia}\ and\ \citenamefont
  {Alexander}(2017)}]{khoromskaia2017vortex}%
  \BibitemOpen
  \bibfield  {author} {\bibinfo {author} {\bibfnamefont {D.}~\bibnamefont
  {Khoromskaia}}\ and\ \bibinfo {author} {\bibfnamefont {G.~P.}\ \bibnamefont
  {Alexander}},\ }\href {\doibase 10.1088/1367-2630/aa89aa} {\bibfield
  {journal} {\bibinfo  {journal} {New J. Phys.}\ }\textbf {\bibinfo {volume}
  {19}},\ \bibinfo {pages} {103043} (\bibinfo {year} {2017})}\BibitemShut
  {NoStop}%
\bibitem [{\citenamefont {Si{\'e}fert}\ \emph {et~al.}(2019)\citenamefont
  {Si{\'e}fert}, \citenamefont {Reyssat}, \citenamefont {Bico},\ and\
  \citenamefont {Roman}}]{Siefert:2019bi}%
  \BibitemOpen
  \bibfield  {author} {\bibinfo {author} {\bibfnamefont {E.}~\bibnamefont
  {Si{\'e}fert}}, \bibinfo {author} {\bibfnamefont {E.}~\bibnamefont
  {Reyssat}}, \bibinfo {author} {\bibfnamefont {J.}~\bibnamefont {Bico}}, \
  and\ \bibinfo {author} {\bibfnamefont {B.}~\bibnamefont {Roman}},\ }\href
  {https://doi.org/10.1038/s41563-018-0219-x} {\bibfield  {journal} {\bibinfo
  {journal} {Nat. Mater.}\ }\textbf {\bibinfo {volume} {18}},\ \bibinfo {pages}
  {24} (\bibinfo {year} {2019})}\BibitemShut {NoStop}%
\end{thebibliography}%

\end{document}